\documentclass[twocolumn,showpacs,preprintnumbers,amsmath,amssymb,superscriptaddress]{revtex4}

\usepackage{graphicx}
\usepackage{dcolumn}
\usepackage{bm}

\begin{document}

\title{Temperature and Magnetic Field Dependencies of Condon Domain Phase\\in Lifschitz-Kosevich-Shoenberg Approximation }

\author{Nathan Logoboy}

\email{logoboy@phys.huji.ac.il}

\affiliation{Grenoble High Magnetic Field Laboratory, MPI-FKF and
CNRS P.O. 166X, F-38042 Grenoble Cedex 9, France}

\affiliation {The Racah Institute of Physics, The Hebrew University
of Jerusalem, 91904 Jerusalem, Israel}

\author{Walter Joss}
\affiliation{Grenoble High Magnetic Field Laboratory, MPI-FKF and
CNRS P.O. 166X, F-38042 Grenoble Cedex 9, France}

\affiliation {Universit$\acute{e}$ Joseph Fourier, B.P. 53, F-38041
Grenoble Cedex 9, France}

\date{\today}

\begin{abstract}
The temperature and magnetic field behavior of non-uniform
diamagnetic phase of strongly correlated electron gas at the
conditions of dHvA effect is analyzed. It is shown, that in the
framework of Lifschitz-Kosevich-Shoenberg approximation the magnetic
induction splitting, as well as the range of existence of Condon
domains, are characterized by strong dependencies on temperature,
magnetic field and impurities of the sample.
\end{abstract}

\maketitle

\section{\label{sec:Introduction}Introduction}

Oscillations of the thermodynamic quantities  of electron gas in
magnetic field are the result of the oscillations of the density of
states when successive Landau levels sweep through the Fermi level
\cite{Lifshitz}. At high magnetic field and low temperature the
instability of strongly correlated electron gas due to magnetic
interaction between electrons results in diamagnetic phase
transition (DPT) \cite{Shoenberg} with formation of Condon domain
(CD) structure, which is extensively studied, both theoretically and
experimentally \cite{Condon}-\cite{Gordon Soft Mode}. Magnetic
interaction of strongly correlated electron gas results in
non-linear dependence of local diamagnetic moments on external
magnetic field and temperature \cite{Logoboy1} and gives rise such
an exotic phenomenon as diamagnetic hysteresis \cite{Kramer2},
\cite{Logoboy5}.

Recent progress in experiments on observation of CD structure in Ag
\cite{Kramer1} and Be \cite{Kramer2} provides a natural stimulus
towards a more detailed understanding of the properties of strongly
correlated electron gas in the conditions of dHvA effect.

Although the theoretical aspects of the CD formation have been
recently reviewed \cite{Gordon}, some important questions concerning
the DPT and non-uniform phase of strongly correlated electron gas
are still open. In particular, the detailed theoretical
investigation of the influence of the temperature $T$, magnetic
field $\mu_{0}H$ and impurities of the sample on such important
characteristics of the non-uniform diamagnetic phase, as the range
of existence of the CD structure and magnetic induction splitting,
is still lacking.

We present the theoretical investigation of the temperature and
magnetic field behavior of the CD phase in the framework of
Lifshitz-Kosevich-Shoenberg (LKS) approximation \cite{Lifshitz},
\cite{Shoenberg} and compare our results with available data on
investigation of CDs in Ag \cite{Condon_Walstedt},\cite{Kramer1}.

\section{\label{sec:Model}Model}

The oscillator part of the thermodynamic potential density in the
LKS approximation \cite{Shoenberg} can be written in reduced form
with taking into account the shape sample effects:

\begin{equation} \label{eq:free energy density}
\Omega =\frac{1} {4 \pi k^{2}} \left [ a \cos {b}+\frac{1}{2}a^{2}(1-n) \sin^{2} {b} \right ], \\
\end{equation}
where $b=k(B-\mu_{0}H)=k[h_{ex}+4 \pi (1-n)M]=x+(1-n)y$, $\mu_{0}H$
is the magnetic field inside the material $k=2 \pi
F/(\mu_{0}H)^{2}$, $F$ is the fundamental oscillation frequency),
$h_{ex}=H_{ex}-H$ is the small increment of the magnetic field $H$
and the external magnetic field $H_{ex}$, $n$ is the demagnetization
factor.
\begin{figure}[b]
  \includegraphics[width=0.4\textwidth]{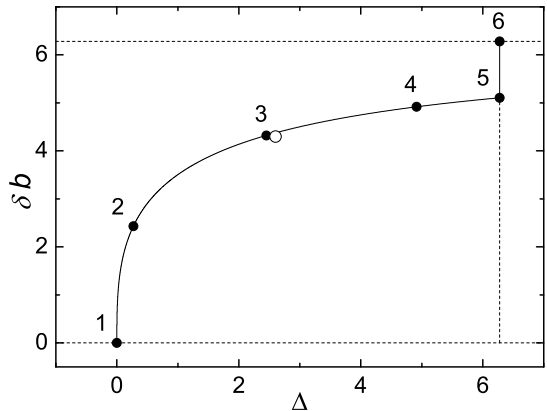}
\caption{ The magnetic induction splitting $\delta b$ as a function
of the range of existence of non-uniform phase $\Delta$. Closed
circles show the different values of reduced amplitude of dHvA
oscillations $a$: $1$ - $a=1$, $2$ - $a=1.3$, $3$ - $a=2.6$, $4$ -
$a=3.9$, $5$ - $a=4.6$, $2$ - $a \to \infty$. Both variables $0 \le
\delta b \le 2\pi$ and $0 \le \Delta \le 2\pi$ are restricted by the
period of the oscillations ($2 \pi$ in reduced units), which is
shown by dash lines. The straight line between points $5$ and $6$
corresponds to slow increase of induction splitting $\delta b$ with
increase of $a$ in the interval $a \in [4.6, \infty)$, when $\Delta$
has reached its maximum value $\Delta =2 \pi$. The open circle
corresponds to the data \cite{Condon_Walstedt}. }
\label{Splitting-Range}
\end{figure}
The validity of thermodynamic potential density (\ref{eq:free energy
density}) is restricted by applicability to homogeneous phase only,
where the conception of demagnetization coefficient $n$ is justified
\cite{Logoboy1}. In the conditions of strong magnetic interaction,
when

\begin{equation} \label{eq:CDCondition}
a \left ( \mu_{0}H, T, T_{D}\right )\ge 1, \\
\end{equation}
a state of lower thermodynamic potential is achieved over part of
dHvA oscillation cycle by the formation of CD structure. The usually
observed diamagnetic domain structure is of stripe-domain type
\cite{Shoenberg},\cite{Kramer1}. In the domain state, when the
reduced amplitude of dHvA oscillations satisfies to the condition
(\ref{eq:CDCondition}), the demagnetization factor $n$ is replaced
by the coefficient $\alpha$ \cite{Logoboy1}, which depends on
magnetic field, temperature, impurities and geometry of the sample.

The equation $a(\mu_{0}H, T, T_{D})= 1$ defines the critical surface
in three dimensions $\mu_{0}H-T-T_{D}$ . Above this surface the
uniform diamagnetic phase exists, but below it, the CD phase appears
in the part of the period of dHvA oscillations.

In the non-uniform phase, when the condition (\ref{eq:CDCondition})
is fulfilled, the metastable states, formed by magnetic field $h$,
are determined by

\begin{equation} \label{eq:Metastable States}
\frac {\partial \Omega}{\partial x}=0, \quad \frac{\partial^{2}\Omega} {\partial x^{2}} >0, \\
\end{equation}
identifying two local minima of the thermodynamic potential
(\ref{eq:free energy density}). The diamagnetic moments $y_{\pm}$
corresponding to these minima, being the functions of magnetic field
increment $x$ and reduced amplitude of the dHvA oscillations
$a=a(\mu_{0}H,T,T_{D})$ \cite{Logoboy1}, are characterized by the
strong dependence on magnetic field, temperature and impurity of the
sample and contribute to the magnetic induction splitting $\delta
b=y_{+}-y_{-}$ between two adjacent domains. The average magnetic
induction splitting can be defined as the magnetic induction
splitting at the center of the period of dHvA oscillations:

\begin{equation} \label{eq:Splitting}
\delta b(x;a)\approx \delta b(0;a)=y_{+}(0;a)-y_{-}(0;a). \\
\end{equation}
In this case the magnetic induction splitting $\delta b$ is
calculated due to

\begin{equation} \label{eq:Equation for Splitting}
\delta b=2a \sin{ \frac{\delta b}{2}}. \\
\end{equation}
In limit $a \to 1+0^{+}$, e.g. in the nearest vicinity of the point
of DPT, from the Eq.~(\ref{eq:Equation for Splitting}) one can
obtain the expression for magnetization $y_{\pm}=\pm
\sqrt{[6(1-1/a)]}$ \cite{Gordon}. According to Eq.~(\ref{eq:Equation
for Splitting}) magnetic induction splitting $\delta b =\delta b(a)$
has strong dependencies on magnetic field $\mu_{0}H$, temperature
$T$ and Dingle temperature $T_{D}$ through the reduced amplitude of
dHvA oscillations $a=a(\mu_{0}H,T,T_{D})$.

The range of existence of the CD structure $\Delta$ in every period
of dHvA oscillations can be estimated by consideration of the
instability points, or inflection points, where the next two
conditions must simultaneously hold

\begin{equation} \label{eq:Inflection Points}
\frac {\partial \Omega}{\partial x}=0, \quad \frac{\partial^{2}\Omega} {\partial x^{2}} =0. \\
\end{equation}
Using the Eqs.~(\ref{eq:Inflection Points}) one can obtain

\begin{equation} \label{eq:Range}
\Delta =2(\sqrt{a^{2}-1}-\cos^{-1}{\frac{1}{a}}). \\
\end{equation}
From Eq.~(\ref{eq:Range}) it follows that the range of existence of
the CD structure is also dependent on magnetic field $\mu_{0}H$,
temperature $T$ and Dingle temperature $T_{D}$.

For proper calculation of the reduced amplitude of the dHvA
oscillations $a$ and investigation of the  temperature and magnetic
field characteristics of CD phase, the correct topology of Fermi
surface has to be taken into account.

In case of ellipsoidal Fermi surface, the temperature and field
dependence of the reduced amplitude of dHvA oscillations $a$ is
defined by \cite{Shoenberg},\cite{Gordon}

\begin{equation} \label{eq:Reduced Amplitude}
a=a_{0}(\mu_{0}H)\frac{\lambda(\mu_{0}H,T)}{\sinh{\lambda(\mu_{0}H,T)}}\exp{[-\lambda(\mu_{0}H,T_{D})]}, \\
\end{equation}
where

\begin{equation} \label{eq:Lambda}
\lambda(\mu_{0}H,T)=\frac{2\pi^{2}k_{B}m_{c}cT}{e \hbar \mu_{0}H}, \\
\end{equation}
$m_{c}$ is the cyclotron mass, $k_{B}$ is the Boltzmann constant,
$e$ is the absolute value of the electron charge, $c$ is the light
velocity, $\hbar$ is the Planck constant, and $T_{D}=\hbar /2 \pi
k_{B}\tau$ is the Dingle temperature, which is inversely
proportional to the scattering lifetime $\tau$ of the conduction
electrons.

The limiting amplitude

\begin{equation} \label{eq:Limiting Amplitude}
a_{0}\equiv a(T \to 0,T_{D} \to 0)=\big(\frac{H}{H_{m}}\big)^{3/2} \\
\end{equation}
is the combination of temperature-independent factors in the
Lifschitz-Kosevich formula \cite{Lifshitz}, and

\begin{equation} \label{eq:Limiting Field}
\mu_{0}H_{m}=(10.4\eta\epsilon ^{2}_{F})^{2/3} \\
\end{equation}
is the limiting field \cite{Gordon} above which DPT does not occur
at any temperature $T$, $\epsilon_{F}$ is Fermi energy in $eV$,
$\eta=m_{c}/m$ and $m$ electron mass. The validity of the LKS
approximation and, consequently, the expression for reduced
amplitude of the dHvA oscillations $a$ (\ref{eq:Reduced Amplitude})
is restricted by the application to the spherical (or almost
spherical) Fermi surface sheets, which is the case of noble metals
\cite{Shoenberg}. The recent experimental data on observation of CD
structure in Ag \cite{Kramer2}, \cite{Kramer3} by measurement of the
amplitude of the third harmonic of the $ac$ susceptibility reveals
the possibility to construct the phase diagram. As it was expected,
the measured phase diagrams for Ag \cite{Kramer3} are in a good
agreement with the calculated in the framework of LKS formalism
phase diagrams, justifying the applicability of the
Eq.~(\ref{eq:Reduced Amplitude}) for belly oscillations in Ag.
\begin{figure}
  \includegraphics[width=0.4\textwidth]{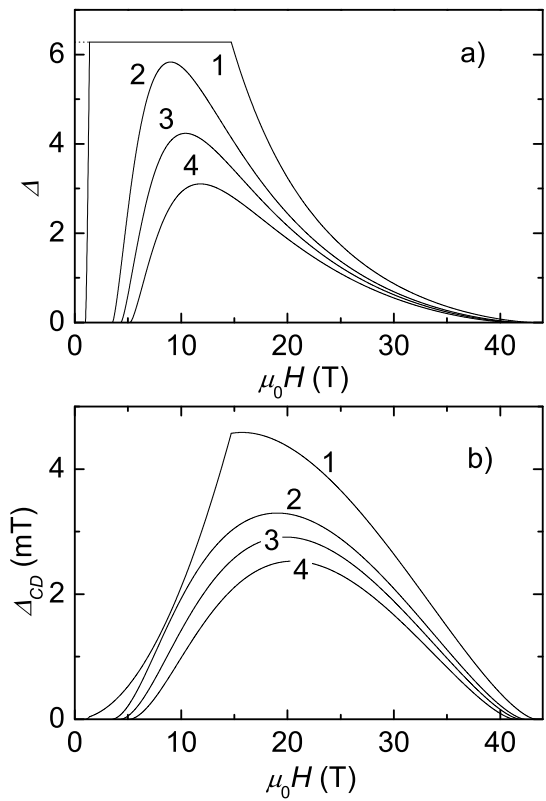}
\caption{\ Magnetic field dependence of the reduced range of
existence of non-uniform phase $\Delta$ (a) and corresponding
measured range $\Delta_{CD}$ (a) in silver for $T_{D}=0.1 K$ and
four different temperatures: $1$ - $T=1 K$, $2$ - $T=1.5 K$, $3$ -
$T=1.75 K$, $4$ - $T=2 K$. The straight horizontal line at (a)
corresponds to the maximum value of $\Delta =2 \pi$. This line
transforms into second power of magnetic field dependence of
$\Delta_{CD}=\Delta/k$ at the same interval of magnetic field (b).
At $T=1 K$ $\Delta_{CD}$ reaches its maximum value at $
\mu_{0}H\approx 15 T$.} \label{Range H T}
\end{figure}

\section{\label{sec:Results and Discussions}Results and Discussions}

When the condition (\ref{eq:CDCondition}) is fulfilled, the sample
is divided into domains with up and down magnetization. Both the
magnetic induction splitting defined in explicit form by the
Eq.~(\ref{eq:Equation for Splitting}) and the range of existence of
CDs Eq.~(\ref{eq:Range}) are the functions of reduced amplitude of
the dHvA oscillations and cannot be considered as independent
functions at wide range of applied magnetic field and temperature.
Mathematically, the Eqs.~(\ref{eq:CDCondition}),(\ref{eq:Equation
for Splitting}) define the function $\delta b=\delta b(\Delta)$ in
parametric form with the parameter $a \in [1,~4.6]$
(Fig.~\ref{Splitting-Range}). At $a=4.6$ the reduced range of
existence of CD structure spreads through all period of the dHvA
oscillations and remains constant at further increase of $a$.

\begin{figure}[b]
  \includegraphics[width=0.4\textwidth]{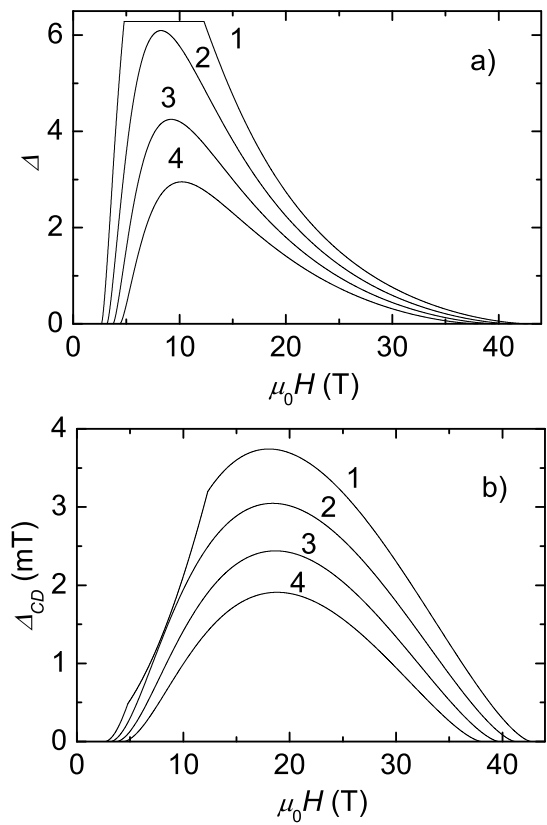}
\caption{\ Magnetic field dependence of the reduced range of
existence of non-uniform phase $\Delta$ (a) and corresponding
measured range $\Delta_{CD}$ (a) in silver for $T=1.2 K$ and four
different Dingle temperatures: $1$ - $T_{D}=0.1 K$, $2$ -
$T_{D}=0.25 K$, $3$ - $T_{D}=0.4 K$, $4$ - $T_{D}=0.55 K$. Similar
to the Fig.~\ref{Range H T} the straight horizontal line at (a)
corresponds to the maximum value of $\Delta =2 \pi$. This line
transforms into second power of magnetic field dependence of
$\Delta_{CD}=\Delta/k$ at the same interval of magnetic field in
(b). $\Delta_{CD}$ reaches its maximum value at $ \mu_{0}H\approx 20
T$.} \label{Range H TD}
\end{figure}
The results of numerical calculations (Fig.~\ref{Splitting-Range})
show rapid increase of magnetic induction $\delta b$ with increase
of reduced amplitude $a$ in the vicinity of the point of DPT with
slowing down of this increase at $a \in [2,~4.6]$. Contrary to it,
the range of existence of the CD structure $\Delta$ is characterized
by slow increase near the DPT point and rapid increase at the
biggest values of $a$ till its full value $\Delta=2 \pi$ at $a=4.6$.
Thus, $\delta b=2.43>> \Delta=0.27$ at $a=1.3$. At the interval $a
\in [4.6, \infty]$ the function $\delta b$ increases slowly with
increase of $a$, while the range of existence of CDs remains
constant $\Delta=2\pi$. In famous experiment on observation of CD
structure in Ag by NMR \cite{Condon_Walstedt} the value of $a$ was
found to be about $2.6$, which is consistent with the value of $a$,
calculated from Eq.~(\ref{eq:Reduced Amplitude}) at the conditions
of the experiment: $\mu_{0}H=9 T$, $T=1.4 K$ and $T_{D}=0.8 K$.
Using the reported values of the dHvA period $\Delta H \approx 17
G$, $\delta B/ \mu_{0}\approx 12 G$ and evaluating the range of
existence of the CD structure as $\Delta_{CD}\approx 7 G$, one can
estimate $\delta b=2\pi \delta B/\mu_{0}\Delta H\approx 4.3$ and
$\Delta =k \Delta_{CD} \approx 2.6$. These values are very close to
$\delta b =4.4$ and $\Delta=2.5$, calculated from
Eqs.~(\ref{eq:Splitting}) and (\ref{eq:Range}).

 Due to the bell-like shape of the diamagnetic phase
diagrams $T=T(\mu_{0}H,T_{D})$ (see, e. g. \cite{Gordon}) there is
one critical temperature $T_{c}$ at given magnetic field and two
critical values of magnetic field $H_{\mp}$ ($H_{-}<H_{+}$) at given
temperature. Another possibility for realization of the DPT is
related to the concentration of impurities in the sample, which
influence on the amplitude of the dHvA oscillations through the
scattering $\tau$ of conduction electron.

Near the point of DPT $a \to 1+0^{+}$ ($T \to T_{c}+0^{-}$, or $H
\to H_{\mp}+0^{\pm}$, or $T_{D} \to T_{D,c}+0^{-}$) the temperature
and magnetic field dependencies of magnetic induction $\delta b$ can
be represented as follows:

\begin{equation}\label{eq:T H Splitting}
\delta b = \left\{ \begin{array}{ll}
  2[6 \lambda_{c}L(\lambda_{c})]^{1/2} \big( \displaystyle \frac {T_{c}-T}{T_{c}}\big)^{1/2}, &T\to T_{c}+0^{-}  \\ \\
  2(6 \lambda_{D,c})^{1/2} \big( \displaystyle \frac {T_{D,c}-T_{D}}{T_{D,c}}\big)^{1/2}, &T_{D}\to T_{D,c}+0^{-}  \\ \\
  2(6\alpha _{\mp})^{1/2}\big( \pm \displaystyle \frac {H-H_{\mp}}{H_{\mp}}\big)^{1/2}, &H\to H_{\mp}+0^{\pm} \\
 \end{array} \right.
\end{equation}
where $L(t)=\coth t-1/t$ is Langevin function and

\begin{equation} \label{eq:Alpha}
\alpha _{\mp}=\pm \big [-\frac{3}{2}+\lambda_{\mp}L(\lambda_{\mp})-\lambda^{(D)}_{\mp} \big]. \\
\end{equation}

Here, $\lambda_{c}=\lambda(\mu_{0}H,T_{c})$,
$\lambda_{D,c}=\lambda(\mu_{0}H,T_{D,c})$,
$\lambda_{\mp}=\lambda(\mu_{0}H_{\mp},T)$,
$\lambda^{(D)}_{\mp}=\lambda(\mu_{0}H_{\mp},T_{D})$. It can be shown
that near the point of DPT the range of existence of CD structure
$\Delta$ is related to the magnetic induction splitting $\delta b$
according to

\begin{equation} \label{eq:Relationship}
\Delta =\frac {1}{36}(\delta b)^{3}. \\
\end{equation}

It follows from Eqs.~(\ref{eq:T H Splitting}) and
(\ref{eq:Relationship}) that the parameters of the CD phase are
characterized by the critical behavior near the DPT. The temperature
and magnetic field dependencies of the order parameter $\delta b$
(\ref{eq:T H Splitting}) can result in existence of soft mode. The
possibility of softening of the orbital magnon mode was studied in
\cite{Gordon Soft Mode}.

\begin{figure}
  \includegraphics[width=0.4\textwidth]{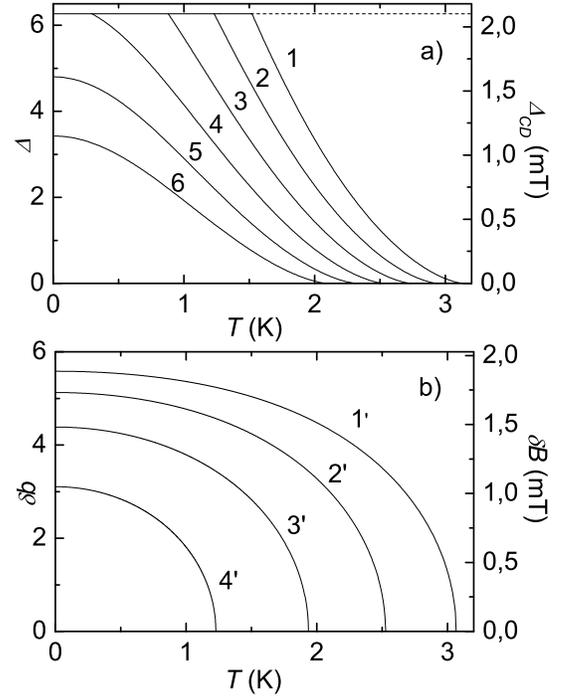}
\caption{(a) Temperature dependence of the range of existence of CD
phase for six different Dingle temperatures: $1$ - $T_{D}=0.1 K$,
$2$ - $T_{D}=0.25 K$, $3$ - $T_{D}=0.4 K$, $4$ - $T_{D}=0.55 K$, $5$
- $T_{D}=0.7 K$ and $6$ - $T_{D}=0.85 K$. (b) Temperature dependence
of the magnetic induction splitting due to CDs for four different
Dingle temperatures: $1'$ - $T_{D}=0.1 K$, $2'$ - $T_{D}=0.5 K$,
$3'$ - $T_{D}=0.9 K$ and $4'$ - $T_{D}=1.3 K$. The strait horizontal
line at (a) corresponds to the maximum value of $\Delta =2
\pi$.}\label{Range T TD}
\end{figure}
In general case the temperature and magnetic field characteristics
of CD structure can be calculated from Eqs.~(\ref{eq:Splitting}) and
(\ref{eq:Range}) with taking into account the Eq.~(\ref{eq:Reduced
Amplitude}) for reduced amplitude of the oscillations. To illustrate
the behavior of the non-uniform diamagnetic phase, we calculated the
temperature and magnetic field dependencies of the range of
existence of CDs $\Delta$ and magnetic induction splitting $\delta
b$ for Ag, where the domains were observed \cite{Kramer1} and the
DPT diagrams were confirmed experimentally \cite{Kramer3}.

The measured, or absolute, values of the range of existence of CD
structure $\Delta_{CD}$ and magnetic induction splitting $\Delta B$
are defined as follows

\begin{eqnarray}
\Delta _{CD}=\frac{\Delta}{k}=\Delta \frac{(\mu_{0}H)^{2}}{2\pi F},\qquad \label{eq:Measured Range}\\
\Delta B=\frac{\delta b}{k}=\delta b \frac{(\mu_{0}H)^{2}}{2\pi F},
\label{eq:Measured Splitting}
\end{eqnarray}

The results of numerical calculation of the temperature and magnetic
field dependencies of the parameters of the CD structure, $\Delta$
Eq.~(\ref{eq:Range}) and $\delta b$ Eq.~(\ref{eq:Equation for
Splitting}), are illustrated in Figs.~\ref{Range H T}-\ref{Splitting
H TD}.

Fig.~\ref{Range H T} shows the magnetic field dependencies of the
range of existence of CD structure in Ag at constant Dingle
temperature $T_{D}=0.1 K$ and several temperatures: $T=1 K$, $T=1.5
K$, $T=1.75 K$, and $T=2 K$. The straight horizontal line in the
graphical representation of the function $\Delta =\Delta(\mu_{0}H)$
at $T_{D}=0.1 K$ (Fig.~\ref{Range H T} (a), curve $1$) corresponds
to the full value of reduced range $\Delta= 2\pi$ with the amplitude
of oscillations $a\ge 4.6$ (compare with the straight line $5$ -
$6$, Fig.~\ref{Splitting-Range}). Corresponding measured, or
absolute, range of non-uniform phase existence $\Delta_{CD}$
(\ref{eq:Measured Range}) is characterized by the second power field
behavior at the same applied field interval (Fig.~\ref{Range H T}).
The function $\Delta_{CD}$ has maximums at the middle range of
applied field $\mu_{0}H \in (15,~25 T)$ and decreases with increase
of the impurity of the sample.

In Fig.~\ref{Range H TD} the range of existence on CDs is shown as a
function of magnetic field for $T=1.2 K$ and four different Dingle
temperatures: $T_{D}=0.1 K$, $T_{D}=0.25 K$, $T_{D}=0.4 K$, and
$T=0.55 K$. The growth of Dingle temperature results in decrease of
the range of the non-uniform phase. Similar to Fig.~\ref{Range H T},
the straight horizontal line in Fig.~\ref{Range H TD}(a) corresponds
to the maximum of reduced range $\Delta$, which results in quadratic
increase of the absolute range of CD existence $\Delta_{CD}$
(\ref{eq:Measured Range}).

The temperature dependencies of the range of existence of CDs
$\Delta$ (\ref{eq:Range}) ($\Delta_{CD}$ (\ref{eq:Measured Range}))
and magnetic induction splitting $\delta b$ (\ref{eq:Equation for
Splitting}) ($\delta B$ (\ref{eq:Measured Splitting})) at
$\mu_{0}H=10 T$ and different Dingle temperatures $T_{D}$ are
illustrated in Fig.~\ref{Range T TD}. At the temperature interval $T
\sim 0 - 1.5 K$ for $T_{D}= 0.1 K$ (curve $1$ (a)) the range of CD
existence $\Delta$ spreads over all period of dHvA oscillations,
reaching its maximum possible value. Near the point of diamagnetic
phase transition, when  $a \to 1+0^{+}$ the parameter of the order
$\delta b$ exhibits critical behavior in accordance with
Eq.~(\ref{eq:T H Splitting}). It results in softening of orbital
mode. The last circumstance is important for experimental
observation of the DPT point and constructing the phase diagrams
$\mu_{0}H-T-T_{D}$ by measuring the temperature dependence of
parameter of order $\Delta B=\Delta B(T)$. The increase of the
Dingle temperature $T_{D}$ results in decrease of the range $\delta$
(Fig.~\ref{Range T TD}(a)) and splitting $\delta b$ (Fig.~\ref{Range
T TD}(b)).

\begin{figure}
  \includegraphics[width=0.4\textwidth]{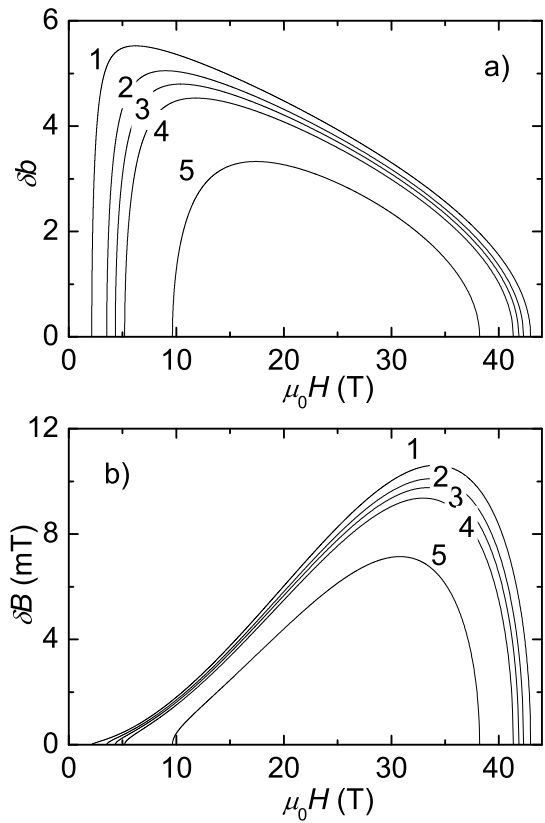}
\caption{\ Magnetic field dependence of the reduced induction
splitting $\delta b$ (a), caused by two adjacent domains, and
corresponding measured value $\delta B$ (b) in silver for $T_{D}=0.1
K$ and five different temperatures: $1$ - $T=1 K$, $2$ - $T=1.5 K$,
$3$ - $T=1.75 K$, $4$ - $T=2 K$, and $5$ - $T=3 K$. The second power
law dependence of the period of the dHvA oscillations on the applied
magnetic field shifts the maximum values of measured characteristics
to the high-field range.} \label{Splitting H T}
\end{figure}

\begin{figure}[b]
  \includegraphics[width=0.4\textwidth]{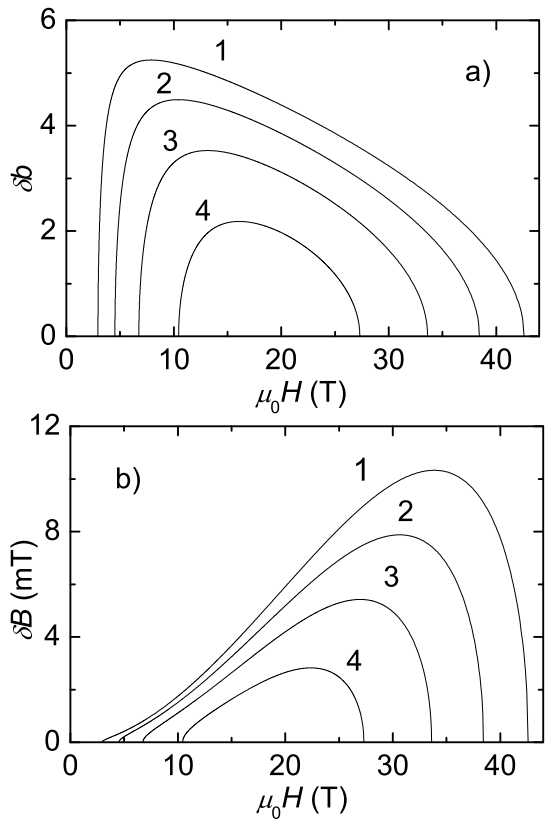}
\caption{\ Magnetic field dependence of the reduced induction
splitting $\delta b$ (a), caused by two adjacent domains, and
corresponding measured value $\delta B$ (b) in silver for $T=1.3 K$
and four different Dingle temperatures: $1$ - $T_{D}=0.1 K$, $2$ -
$T_{D}=0.5 K$, $3$ - $T_{D}=0.9 K$, and $4$ - $T_{D}=1.3 K$. Similar
to Fig.~(\ref{Splitting H T}), the second power law dependence of
the period of the dHvA oscillations on the applied magnetic field
shifts the maximum values of measured characteristics to the
high-field range.} \label{Splitting H TD}
\end{figure}

The magnetic field dependencies of induction splitting $\delta b$
($\delta B$) are shown in Fig.~(\ref{Splitting H T}) for five
different temperatures and in Fig.~\ref{Splitting H TD} for four
different Dingle temperatures. The maximum values of absolute
induction splitting ($\delta B$) are shifted into high-field range
$\mu_{0}H \sim 25 - 40 T$.

The magnetic induction splitting due to CD structure changes the
distribution of magnetic field in vacuum above the surface of the
sample. The possibility of detection of these changes by means of
Hall probe spectroscopy and mapping of CD structure is discussed in
\cite{Logoboy3}, \cite{Logoboy4} and realized in \cite{Kramer1},
\cite{Kramer3}. It was found \cite{Logoboy3}, \cite{Logoboy4} for
the plate-like sample in applied magnetic field along the normal to
the surface $ \boldsymbol H \parallel \boldsymbol Z $ with periodic
domain structure along the $Y$-axis , that the normal component of
magnetic field $H_{n}$ is described by the following periodic
function of $Y$ with the period $2D$ equal to the period of CD
structure :

\begin{equation} \label{eq:Normal Component}
 H_{n} =\frac {\delta B}{\pi \mu_{0}} \tan^{-1}{\frac {\cos{(\pi Y/D)}}{\sinh {(\pi Z/D})}}. \\
\end{equation}

The measured value by method of Hall probes is the maximum splitting
of the normal components of non-uniform magnetic field  above two
neighboring domains $\delta H_{n}$, e. g. the maximum difference
between two values, $H^{\uparrow}_{n}$ and $H^{\downarrow}_{n}$, of
normal component of magnetic field Eq.~(\ref{eq:Normal Component})
caused by magnetic induction splitting $\delta B$ inside the sample

\begin{equation} \label{eq:Field Splitting}
 \delta H_{n} \equiv H^{\uparrow}_{n}-H^{\downarrow}_{n}=\frac{2}{\pi}\frac {\delta B}{\mu_{0}} \tan^{-1}{\frac {1}{\sinh {(\pi Z/D})}}. \\
\end{equation}
The function (\ref{eq:Field Splitting}) is of rapid decrease of $Z$.
Thus, the magnetic field splitting above the sample
Eq.~(\ref{eq:Measured Splitting}) at the distance of one-third of
the period of CD structure is about $0.15\delta B/\mu_{0}$. It
follows from Eq.~(\ref{eq:Field Splitting}) that the measurement of
the magnetic field splitting $\delta H_{n}$ at given distance $Z=d$
{\it above} the surface of the sample can give information about
magnetic induction splitting $\delta B$ {\it inside} the sample due
to CD structure with given period $2D$.

\begin{figure}
  \includegraphics[width=0.4\textwidth]{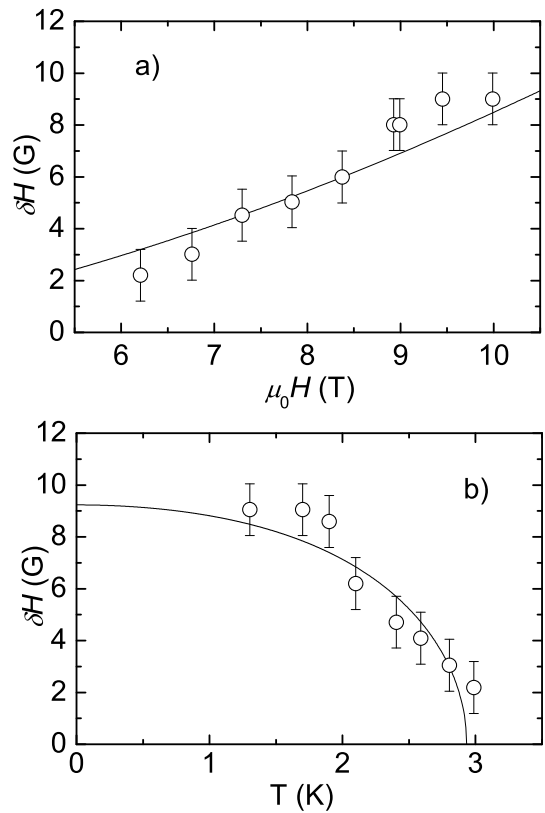}
\caption{Comparison between theory and experiment. (a) Applied
magnetic field dependence (a) and temperature dependence (b) of the
magnetic field splitting $\delta H$ above the surface of the sample,
caused  by the induction splitting $\delta B$ of CD structure. The
solid lines correspond to the magnetic field distribution,
calculated from Eq.~(\ref{eq:Measured Splitting}) in the framework
of LKS approximation at the conditions of the experiment
\cite{Kramer1}: $\mu_{0}H=10 T$ and $T_{D}=0.2 K$. The circles are
referred from \cite{Kramer1}. In calculation of the theoretical
curves the fitting parameter $d/2D=0.1$ is used.} \label{Comparison}
\end{figure}
The Hall probes measurements of magnetic field distribution above
the sample surface \cite{Kramer1}, \cite{Kramer3} were performed on
a high quality single crystal of Ag $2.4\times 1.6\times 1.0
~mm^{3}$. The Dingle temperature was estimated to be $T_{D}=0.2 K$.
At the conditions of the experiment, $\mu_{0}H=10T$ and $T=1.3 K$
\cite{Kramer1}, the magnetic induction splitting, calculated in LKS
approximation with $a$ Eq.~(\ref{eq:Reduced Amplitude}), gives
$\delta B(1.3 K)/\mu_{0}=16.7 G$. These estimations are in agreement
with the earlier famous NMR measurements in Ag
\cite{Condon_Walstedt}, where at the slightly less favorable
conditions for observation of CD structure, e. g. $\mu_{0}H=9 T$,
$T=1.4 K$ and $T_{D}=0.8 K$, the smaller value of induction
splitting in the sample $\delta B/\mu_{0}\approx 12 G$ was reported.
At the conditions of the experiment \cite{Condon_Walstedt}, the
Eq.~(\ref{eq:Measured Splitting}) with taking into account
Eq.~(\ref{eq:Reduced Amplitude}) gives the value of $\delta B(1.4
K)/\mu_{0}=11.7 G$ close to the reported one, justifying the
applicability of the LKS approximation \cite{Shoenberg} for belly
oscillations in Ag.

Correct treatment of the experimental results on measuring of
distribution of magnetic field by Hall probe technic \cite{Kramer1}
can be done with knowledge of the ratio of the distance between the
set of Hall probes and the surface of the sample $d$ to the period
of the domain structure $2D$ (see, e.g. Eq.~(\ref{eq:Field
Splitting})). Unfortunately, the distance between Hall probes and
sample surface was not reported in \cite{Kramer1}, and there are no
accurate data about the characteristic magnetic sizes of the
non-uniform phase, except of the statement, that the domain period
was not smaller than $150\mu m$. The more complex measurements of
the magnetic field distribution at different fixed positions in the
direction perpendicular to the surface of the sample would provide
the lacking information about the period of the CD structure. To
compare the theory with the available experimental results
\cite{Kramer1} we use the fitting parameter $d/2D\approx 0.1$, which
gives reasonable value of $d\approx 15 \mu m$ for $2D=150 \mu m$
within accuracy of the experiment \cite{Kramer4}. At $d\approx 50
\mu m$ which is the reasonable value for the average distance
between unpolished surface and Hall probes, the calculated magnetic
field splitting $\delta H$ Eq.~(\ref{eq:Field Splitting}) is about
$2.5 G$, which is at the edge of accuracy of the experiment
\cite{Kramer1}. It explains the absence of the signal from Hall
probes for rough surface \cite{Kramer1}. The results of comparison
of the temperature and magnetic field dependencies of the measured
magnetic field distribution above the surface of the sample
\cite{Kramer1} with the value of $\delta H_{n}$, calculated from
Eq.~(\ref{eq:Field Splitting}) in LKS approximation
($\ref{eq:Reduced Amplitude}$) are present in
Fig.~(\ref{Comparison}). The reported critical temperature
$T_{c}\approx 3K$ at $\mu_{0}H=10 T$, defined from the temperature
dependence of the measured splitting, is in accordance with the
predicted value $T_{c}=2.9 K$, calculated in LKS approximation for
$T_{D}=0.2 K$ at the same value of $\mu_{0}H=10 T$.

\section{Conclusions}

We investigated theoretically the temperature and magnetic field
characteristics of the CD structure in LKS approximation
\cite{Lifshitz}, \cite{Shoenberg}. We show that the magnetic
induction splitting in non-uniform diamagnetic phase, caused by the
arising instability of strongly correlated electron gas in high
magnetic field and low temperature, and the range of the existence
of this phase depend on temperature, magnetic field and impurity of
the sample. We show that the magnetic induction splitting, e. g. the
order parameter of the DPT, and the range of the existence of this
phase in every dHvA period are dependent on each other functions at
the wide range of magnetic field with possible existence of the
non-uniform diamagnetic phase. We calculated the critical behavior
of the parameters of the non-uniform phase. The theoretical results
are in good agreement with available experimental data on
observation of CDs in Ag. Further experiments on observation of
electron instability and formation of CD structure would provide the
lacking information on characteristic magnetic sizes, e. g. width of
the domains.

\begin{acknowledgments}
We are indebted to V. Egorov, R. Kramer and I. Sheikin for
illuminating discussions. We are also grateful to R.~B.~G.~Kramer
for providing us with his experimental results before publication.
\end{acknowledgments}

\end{document}